\pdfoutput=1

\documentclass[sigconf, nonacm]{acmart}

\begin{document}
\title{Aurora DSQL: Scalable, Multi-Region OLTP}

\author{Marc Brooker}
\affiliation{%
  \institution{Amazon Web Services}
}
\email{mbrooker@amazon.com}

\author{Marc Bowes}
\affiliation{%
  \institution{Amazon Web Services}
}
\email{bowes@amazon.com}

\author{Mike Hershey}
\affiliation{%
  \institution{Amazon Web Services}
}
\email{hersheys@amazon.com}

\author{Zak van der Merwe}
\affiliation{%
  \institution{Amazon Web Services}
}
\email{zakv@amazon.com}

\author{James Morle}
\affiliation{%
  \institution{Amazon Web Services}
}
\email{morlej@amazon.com}

\author{Matthys Strydom}
\affiliation{%
  \institution{Amazon Web Services}
}
\email{strydom@amazon.com}

\begin{abstract}
Aurora DSQL is a serverless SQL database designed for cloud-scale transaction processing with multi-region active-active capabilities. Built on a disaggregated architecture, DSQL separates compute, storage, and transaction coordination into independent, horizontally scalable services. Query processors run in Firecracker MicroVMs executing PostgreSQL-compatible SQL without local state. The system uses multiversion concurrency control with precision timestamps for coordination-free reads and optimistic concurrency control for writes, deferring coordination to commit time through distributed adjudicators and the Journal replication system. This minimizes cross-region latency by requiring coordination only during commits, not individual statements. DSQL enables elastic scaling from zero to millions of transactions per second while providing strong consistency, ACID transactions, and continuous availability during availability zone or region failures.
\end{abstract}

\maketitle

\section{Introduction}
\label{sec:introduction}
Amazon has a long history of building scalable database systems, including Dynamo~\cite{decandia2007} in the 2000s; DynamoDB~\cite{Elhemali2022}, Aurora~\cite{verbitski2017}\footnote{In this paper, when we say Aurora DSQL or DSQL we are referring to the system described in the paper, and when we say Aurora we are referring to Aurora MySQL and Aurora PostgreSQL, a significantly different system}, and Redshift~\cite{gupta2015, armenatzoglou2022} in the early 2010s; and Aurora Serverless~\cite{barnhart2024} and MemoryDB~\cite{taleb2024} in the 2020s. Despite this history, our significant internal adoption of relational databases, and our and our customers scale needs, we did not have a horizontally scalable, scale-out, OLTP-optimized SQL database.

Our product goals for DSQL were based on lessons from more than a decade of learning from AWS database customers. The properties we believed were most important to them were:

\begin{description}
  \item[Serverless] A fully-managed fully-serverless experience (like AWS DynamoDB or Amazon S3), with no infrastructure to manage, monitor, or keep up to date.
  \item[Familiar] An interface that developers already know, allowing them to use the clients, tools, and SQL dialect they're familiar with. Familiarity extends beyond API, to include isolation and transaction semantics, data types, and other SQL concepts.
  \item[Scalable] A database that scales up to millions of transactions per second, and down to zero, reflecting the needs of our customers' dynamic and growing businesses.
  \item[Multi-region active-active] Allowing customers to build active-active architectures than span multiple AWS regions, across continents, with strong consistency, fast failover, and no data loss.
  \item[Strongly consistent] Applications shouldn't need to reason about eventual consistency, and should be able to read and write at any scale with strong consistency.
\end{description}

Our overall goal was to build a relational database system that simplifies the work of application building and operations, freeing builders from worrying about scale, reliability, durability, and even multi-region fault tolerance. A \emph{database of first resort}, which is simple and easy to adopt at low scale, and grows with the application, without adding complexity.

\begin{figure}[htbp]
\centering
\includegraphics[width=0.6\columnwidth]{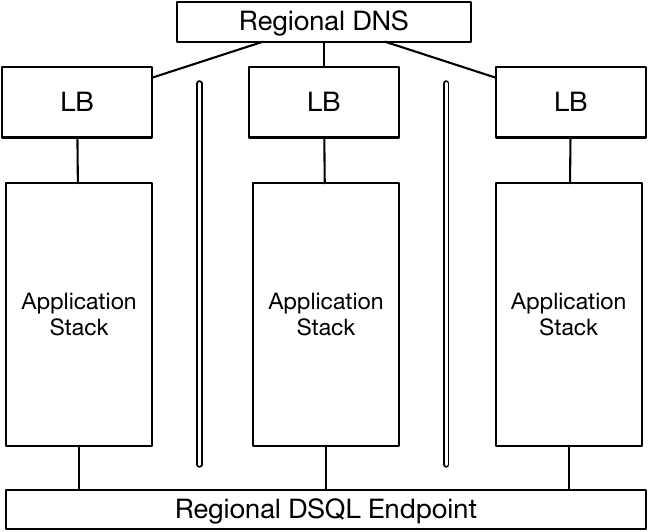}
\caption{Single-region application architecture}
\label{fig:cust_arch_3az}
\end{figure}

Given that our goal is to simplify customers architectures, we'll start by talking about some patterns of scalable, fault-tolerant, cloud-hosted applications that customers build on DSQL. \autoref{fig:cust_arch_3az} shows a single-region architecture, close to what we'd consider the default application architecture for services at Amazon. This architecture, with the application deployed on stateless, auto-scaling compute infrastructure across three datacenters (availability zones, or AZs, in the AWS terminology), backed by a regional DSQL endpoint, allows applications to tolerate the failure of an entire datacenter with no downtime, and has been proven to scale to millions of requests per second and beyond. Responsibility for handling concurrency and durability is handed to the database, and the application fleet is constructed of the required number of stateless, independent, service hosts. Host failures are handled with load balancer (LB) health checks, and datacenter-scale failures are handled with DNS-level health checks.

\begin{figure}[htbp]
\centering
\includegraphics[width=0.8\columnwidth]{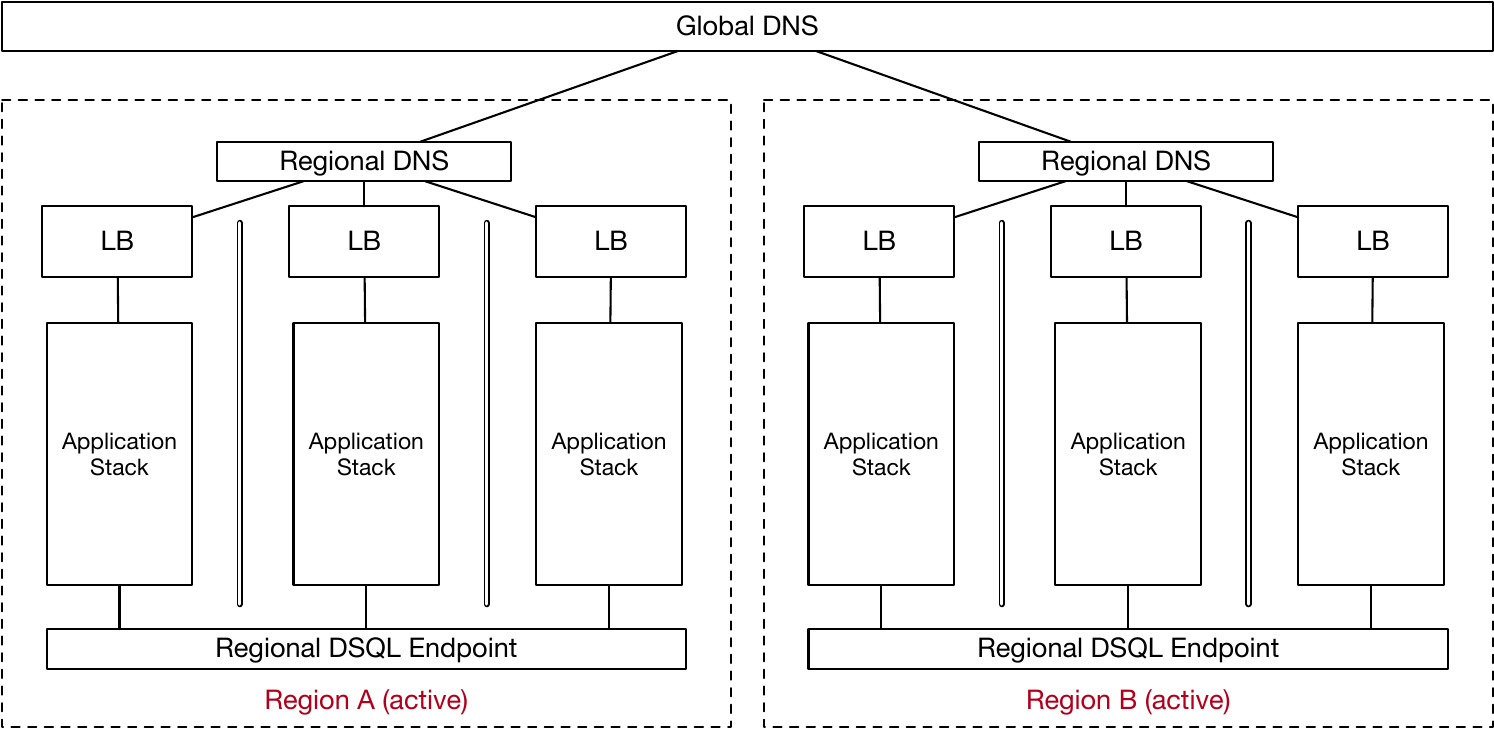}
\caption{Multi-region application architecture}
\label{fig:cust_arch_2region}
\end{figure}

\autoref{fig:cust_arch_2region} extends this architecture to multiple regions (for example us-east-1 in Virginia and us-east-2 in Ohio). In this variant, the application can tolerate the unavailability of an entire region without downtime, and without operator intervention or data loss. From the application programmer's perspective, they continue to write their business logic the same way, and talk to the regional database endpoint. If the application is hosted on serverless compute, they may not even need to manage application infrastructure, while still getting enterprise-ready scale and fault tolerance properties, along with the ability to scale to zero and incur no costs when there is no customer traffic.

Many applications are more complex than this, but we have seen customers and ourselves at Amazon, build global-scale critical businesses on top of such simple architectures. A scalable, active-active, fault-tolerant database is at the core of making such simplicity possible.

In the remainder of this paper, we describe the design and implementation of DSQL, stepping through key decisions. It is not possible to describe a system like this completely in the space provided, so our focus will be on SQL execution, replication, and transactions, leaving details like on-disk storage, encryption, and query optimization to later publications.

\section{Architecture Overview}
\label{sec:architecture}

DSQL's architecture is disaggregated. Multiple independent services, each focused on a small number of well-defined concerns. These components communicate through carefully specified APIs with clear contracts. This approach delivers three crucial properties: we can make independent changes and improvements to each component, we can scale components individually based on workload demands, and we can make different security isolation decisions for each component.

\begin{figure}[htbp]
\centering
\includegraphics[width=0.6\columnwidth]{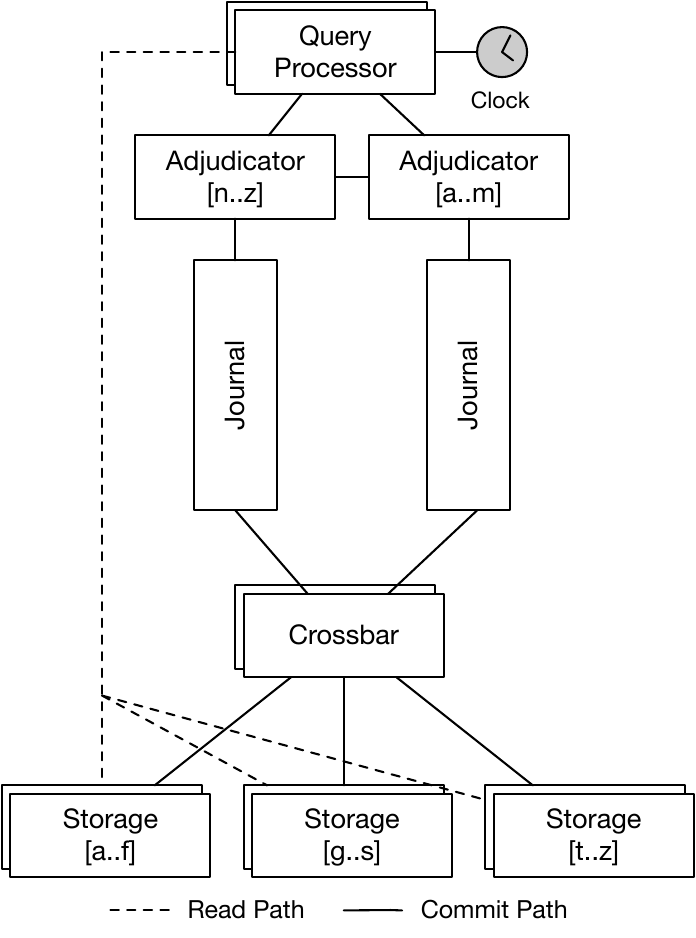}
\caption{Aurora DSQL Architecture Overview}
\label{fig:arch_overview}
\end{figure}

Figure \ref{fig:arch_overview} shows how these components fit together in DSQL's overall architecture. Each component plays a specific role in the system, and crucially, no component exists as a singleton—everything can scale horizontally to meet demand.

Query Processors handle the customer-facing side of the database—they execute SQL queries, return data for reads, buffer writes locally, and coordinate the transaction protocol. When you connect to DSQL and run a query, you're talking to a Query Processor running in a Firecracker MicroVM. As described in \autoref{sec:sql-engine}, Query Processors dynamically scale to meet the customer workload, both vertically (using similar techniques to Aurora Serverless~\cite{barnhart2024}) and horizontally by adding new Query Processors per active connection.

Reads are handled by the query processor and the storage nodes. Storage nodes provide the foundation for data access, handling both table data and indexes. When Query Processors need data, they read it from storage nodes. Each storage node stores a range of data based on a shard key. Query Processors ask for data \emph{as of} a specific timestamp, enabling a coordination-free read path. Storage nodes scale in two dimensions: enough shards to handle all writes to that table, and within each shard, as many read replicas as needed to handle read traffic. The details of how this multiversion concurrency control works and enables fast, consistent reads are covered in \autoref{sec:reads}.

Writes (\texttt{UPDATE}s, \texttt{INSERT}s, etc) are handled locally inside each Query Processor, expanding out to the rest of the write path on transaction commit.

Adjudicators decide whether transactions can commit while maintaining isolation guarantees, and control transaction order and commit time. Adjudicators scale by sharding. Each key in the database belongs to at most one adjudicator at any given time, distributing the conflict detection workload. Adjudicator sharding and storage sharding are unrelated, allowing sharding based on read and write heat to be done independently (for example, a database with low write traffic and high read traffic could have one adjudicator shard, and many storage shards). We explore the transaction protocol and how adjudicators handle write conflicts in detail in \autoref{sec:writes}.

Journal is an internal replication component we use in many systems at AWS, including S3, DynamoDB, and MemoryDB. Each Journal is a durable, ordered, atomic data stream. In the single-region setting, a Journal commit means that data is durable to storage in two or more AZs, while in the cross-region configuration data is durable in two or more AWS regions. These ordered data streams make transactions durable and replicate data between availability zones and regions. When a transaction commits, it gets written to a Journal, making it permanent and ensuring it propagates to all the places it needs to be visible. Journals scale per transaction. Each transaction's writes go atomically to a single journal, but we can have as many journals as we need.

Crossbars take data from Journals, totally ordered on each Journal, and merge-sort them into a total order for each subscriber shard. Then they divide this ordered stream into shards that align with how the storage layer is partitioned. This component scales with the number of storage nodes and handles the complexity of turning the transaction stream into something storage can efficiently consume.

DSQL's disaggregated approach was not only chosen for scalability. It provides significant advantages in durability, fault tolerance, and availability. There's no single point of failure in the system. No single machine, network link, or even entire datacenter failure can bring down a DSQL cluster. Data isn't stored in just one place, and processing isn't tied to specific hardware.

The result is a database that can scale from zero to millions of transactions per second, maintain strong consistency and isolation guarantees, and remain available even when entire regions go offline. Each component focuses on what it does best, and the clean interfaces between them let us optimize each piece independently while maintaining the overall system's correctness and performance guarantees. Concretely, each component exposes a narrow interface: query processors accept SQL and produce reads against storage and write sets for adjudicators; adjudicators accept write sets and return commit decisions; Journals accept committed transactions and produce ordered streams; and crossbars consume Journal streams and produce per-shard change feeds for storage.

\subsection{Multi-tenancy and Security}
DSQL is a multi-tenant system. Each physical machine in the system is typically handling traffic for thousands of customer workloads. Where we use the word \emph{server} in this paper (for example when referring to \emph{storage servers}) we mean a logical construct, of which there are many on any given physical machine. Multi-tenancy (the ability to run many uncorrelated workloads on the same hardware) and soft allocation (the ability to allocate CPU and memory to workloads on demand) are key to DSQL's economics, as they are to AWS Lambda~\cite{agache2020} and Aurora Serverless~\cite{barnhart2024}. Packing multiple workloads reduces the peak-to-average ratio of load on each physical machine (approximately at the rate of $\sqrt{\mathrm{loads}}$). Given that resource cost typically scales with peak, and revenue scales with average, the economic benefit is obvious.

There is also a customer benefit. In \autoref{sec:heat} we talk about how DSQL handles long-term changes in heat through sharding and replication, but soft allocation gives us a powerful tool to handle short-term changes by allocating more or less resources locally as workloads change shape.

Having multiple tenants on a physical machine does raise the question of how we isolate customers from each other from a security and performance perspective. This is a deep topic that requires its own paper, but at a high level we use process isolation in cases where we control the code and the implementation is in a memory safe language, and VM isolation using Firecracker where the code is open source or written in an unsafe language (such as with the PostgreSQL engine, see \autoref{sec:sql-engine}).

\subsection{Optimizing for Coordination}\label{sec:coord}
A guiding principle of DSQL's architecture is to minimize coordination between components, and strongly avoiding coordination that is not constant time with regards to scale. Minimizing coordination helps with scalability and availability, but the primary concern is optimizing for latency. Reads, as described in further detail in \autoref{sec:reads}, happen between the client, a query processor, and one or more storage replicas, typically within the same availability zone and always within the same region. High quality clocks and physical time avoid the need to coordinate with other readers or writers. Writes, and the read-modify-writes ubiquitous in OLTP SQL, also happen within the same availability zone and region, and involve the same components.

Coordination is required for concurrency control, consistency, and durability at transaction commit time. In the multi-region setting, the two tasks requiring cross-region communication (coordination for consistency and isolation, and replication for durability) can be performed in a single round of communication. In the single-region setting, the same is true of in-region cross-AZ communication.

To achieve this minimal communication, DSQL uses a Optimistic Concurrency Control (OCC), following a variant of Kung and Robinson's classic scheme~\cite{Kung1981}. We avoid the commonly-cited downsides of OCC in two ways: using MVCC to ensure that all reads happen from a consistent snapshot, never requiring that transactions are aborted because they have read old data; and by picking snapshot isolation rather than serializability (we revisit our reasoning for this decision in \autoref{sec:snap_isol}), removing the need to abort on read-write conflicts. By contrast, a pessimistic approach, such as that taken by Spanner~\cite{Corbett2012} and CockroachDB~\cite{Taft2020}, requires continuous coordination during transaction execution (typically per statement) to maintain distributed lock state.

Another practical advantage of OCC is preventing clients from ever being able to block other clients. At scale, we have found that performance \emph{cross talk} between clients is a significant contributor to tail latency in apps backed by pessimistic relational databases (for example clients which pause for garbage collection while holding a write lock). Operational experience at Amazon has shown that contention on locks, and retries on failure to acquire locks, are frequent contributors to system outages, and metastable failures which drive long recovery times (and similar results have been reported in the literature~\cite{Isaacs2025, Huang2022}). Another frequent contributor to instability, and tail latency, is clients which get stuck while holding locks, and we've even seen multiple post mortems for cases where operators went out to lunch while holding locks!

\subsection{Multi-Region Architecture}

\begin{figure}[htbp]
\centering
\includegraphics[width=0.8\columnwidth]{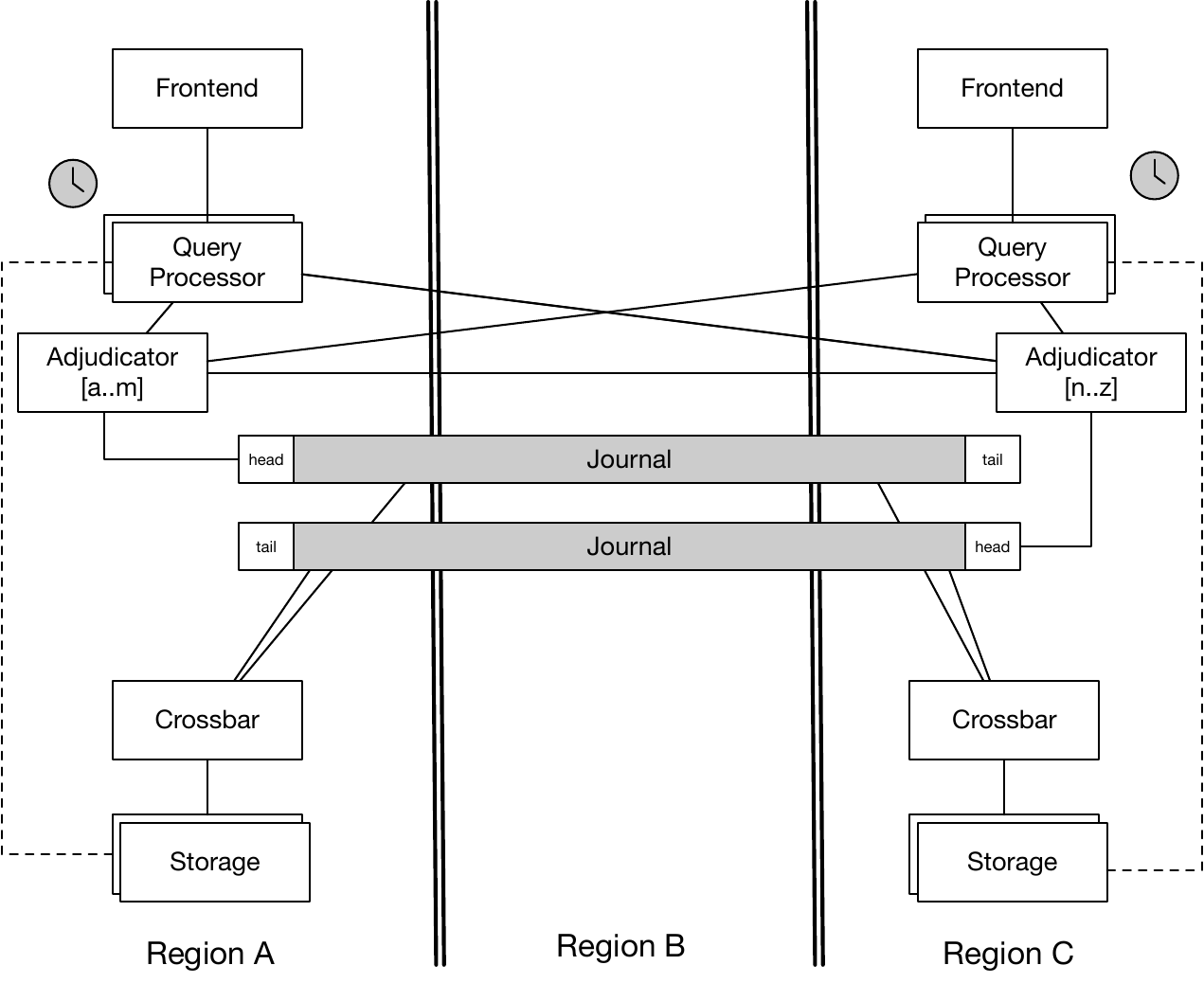}
\caption{Aurora DSQL Architecture Overview}
\label{fig:arch_overview_mr}
\end{figure}

DSQL's multi-region architecture, shown in Figure \ref{fig:arch_overview_mr}, extends the single-region design to provide cross-region durability and availability while minimizing coordination overhead. The fundamental components remain the same, but their behavior and interactions change in important ways.

The read path is nearly unchanged from single-region operation. Read-only transactions and the read portions of read-write transactions are served from storage in the region where the Query Processor runs. The only difference is that storage nodes must wait to see that journals from all regions have advanced past the transaction's read timestamp before serving data, ensuring consistent reads even after region failures.

The write path sees the most significant changes. While writes still buffer locally in Query Processors and commit through Adjudicators, the Journal now replicates transaction data across multiple regions. In the multi-region configuration, Journal commits ensure data durability in two or more AWS regions rather than just two or more availability zones. This cross-region replication happens as part of the commit protocol, requiring only a single round of cross-region communication to achieve both consistency and durability.

Adjudicator placement becomes a control plane optimization problem in multi-region deployments. Each adjudicator shard is placed in a specific region to minimize cross-region traffic based on workload patterns. For failover architectures with one active region, this optimization is straightforward. For active-active workloads with keys spread across regions, optimal placement may not always be possible.

In keeping with our desire to offer customers strong consistency, DSQL chooses consistency over availability when necessary (choosing PCEC in Abadi's PACELC taxonomy~\cite{abadi2012}). However, DSQL remains both strongly consistent and available to clients on the majority side of a network partition. For example, if Region C in Figure \ref{fig:arch_overview_mr} were to fail, DSQL would remain consistent and available to clients in Region A, and keep data durable in Region B.

\subsection{Snapshot Isolation and Strong Consistency}\label{sec:snap_isol}
In DSQL, we picked strong snapshot isolation (i.e. snapshot isolation plus linearizability~\cite{Crooks2017, Daudjee2006}) as our default (and currently only) supported isolation level. This choice has several advantages. First, it is familiar to developers, being equivalent to PostgreSQL's \texttt{REPEATABLE READ} isolation level. Second, it means that transactions in DSQL only abort on write-write conflicts (i.e. only when other concurrent transactions have written the same set of keys), rather than on read-write conflicts as would be required for serializability (i.e. when other transactions have written the keys the committing transaction has read). In OLTP SQL, reads are significantly more common than writes, for the simple reason that most writes (all \texttt{UPDATES}, \texttt{INSERTS} with unique indexes, etc) are also reads. This lower level of isolation means significantly lower abort rates for common transaction patterns.

The downside of snapshot isolation is the introduction of write skew anomalies~\cite{Berenson1995}. Practically, the most important effect of write skew is limiting which business constraints can be enforced by transactions. Working with customers, we have found that giving them tools (such as schema designs that force write-write conflicts for business logic violations, and \texttt{FOR UPDATE}) to avoid the downsides of write skew has enabled the development of correct applications at the snapshot isolation level. We also know from talking to Aurora customers that few choose \texttt{SERIALIZABLE} for their production applications, with most opting for \texttt{REPEATABLE READ} or even \texttt{READ COMMITTED}. DSQL does not currently support \texttt{READ COMMITTED}. This is a simplification enabled by snapshot isolation: because all reads within a transaction see a fixed snapshot at $\tau_{start}$, there is no need for mechanisms like PostgreSQL's EvalPlanQual, which re-evaluates row visibility mid-statement when concurrent transactions commit. Avoiding this complexity reduces the surface area for subtle concurrency bugs in both the database and in applications.

\subsection{Impact of Architectural Properties on Quantitative Performance}
The reduction of round-trip times described in \autoref{sec:coord} is significant for continent-scale systems. In one week in 2022, the network round-trip-time (RTT) between us-east-1 (in Virginia, USA) and us-west-2 (in Oregon, USA) had a mean of $62.17ms$, p50 of $62.4ms$, p99 of $63.30ms$ and p99.99 of $64.99ms$. In a DSQL cluster spanning these two regions, this latency would only be incurred once per transaction. The reality is even better. Journal only needs to commit to two regions of three. So, in a typical multi-region setup spanning us-east-1, us-west-2, and us-east-2 (in Ohio, USA), the commit time that transactions experience would be the time between their client region and the closest second region (for example, the RTT between us-east-1 and us-east-2 has a p50 of $11.5ms$ and p99 of $14.4ms$).

\begin{figure}[htbp]
\centering
\includegraphics[width=0.8\columnwidth]{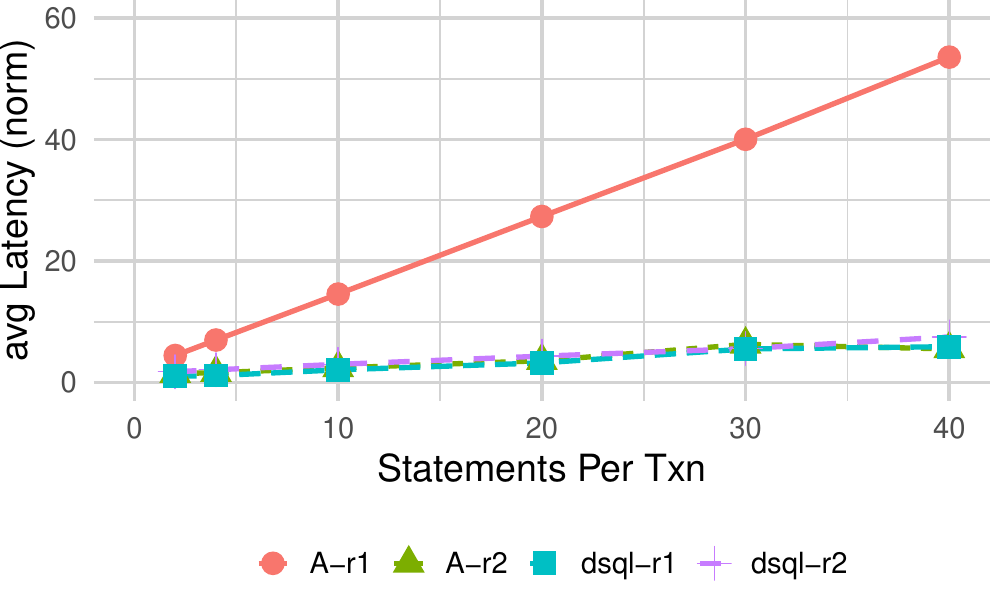}
\caption{Normalized transaction latency comparison between DSQL and Competitor A.}
\label{fig:scale_data}
\end{figure}

\autoref{fig:scale_data} shows the effect of this reduction of round-trip times, comparing DSQL to competitor A, for a multi-region setup with clients in two regions (\emph{r1} and \emph{r2}) across transactions with increasing numbers of \texttt{UPDATE} statements. For competitor A's pessimistic locking-based design, this shows the linear effect of the additional round-trips required to maintain lock state. For DSQL, it shows that latency is the same in both regions.

\begin{figure}[htbp]
\centering
\includegraphics[width=0.9\columnwidth]{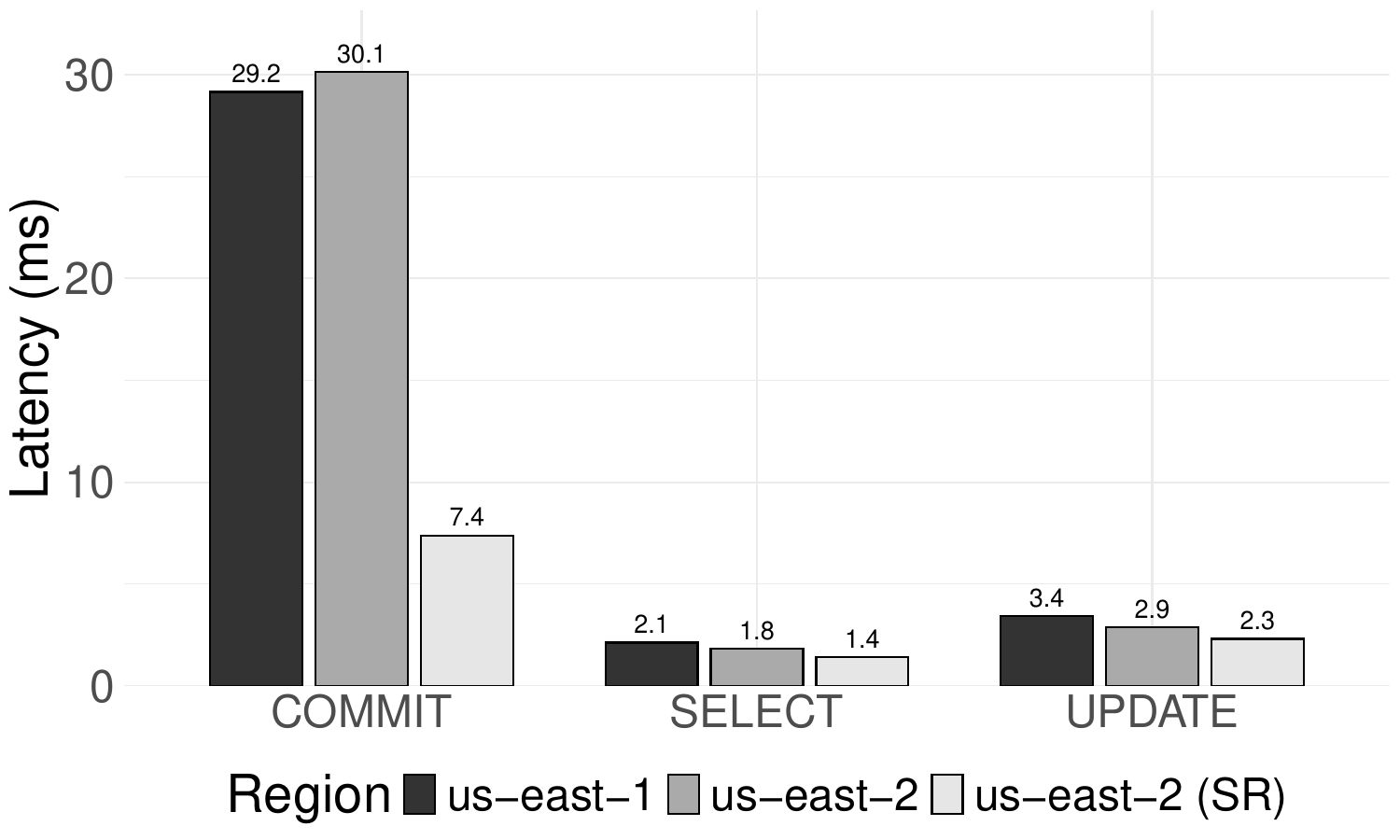}
\caption{DSQL 99th percentile latency for \texttt{COMMIT}, \texttt{SELECT}, and \texttt{UPDATE} operations for a two-region DSQL setup, and a single-region setup.}
\label{fig:low_level_latency}
\end{figure}

\autoref{fig:low_level_latency} zooms in on DSQL's 99th percentile (p99) latency results on a similar microbenchmark, for a cluster spanning \emph{us-east-1} and \emph{us-east-2}, with a witness in \emph{us-west-2}. Clients in each region experience in-AZ read latency ($\approx 2ms$ p99) for primary key \texttt{SELECT} statements, and for \texttt{UPDATE} statements ($\approx 3ms$ p99). Notably, both of these latencies are significantly lower than one network round trip ($14.4ms$ p99) between the pair primary regions, despite offering strong consistency. \texttt{COMMIT} latency is $\approx 30ms$ for the multi-region setup and $7.4ms$ for the single-region setup (note that $\approx 30ms$ is faster than the round-trip to us-west-2 of $63ms$ p99, showing the effect of quorum commit). In the two-regions setup, once the \texttt{COMMIT} is complete, data is durable and available for strongly consistent readers in two regions.

\begin{figure}[htbp]
\centering
\includegraphics[width=0.9\columnwidth]{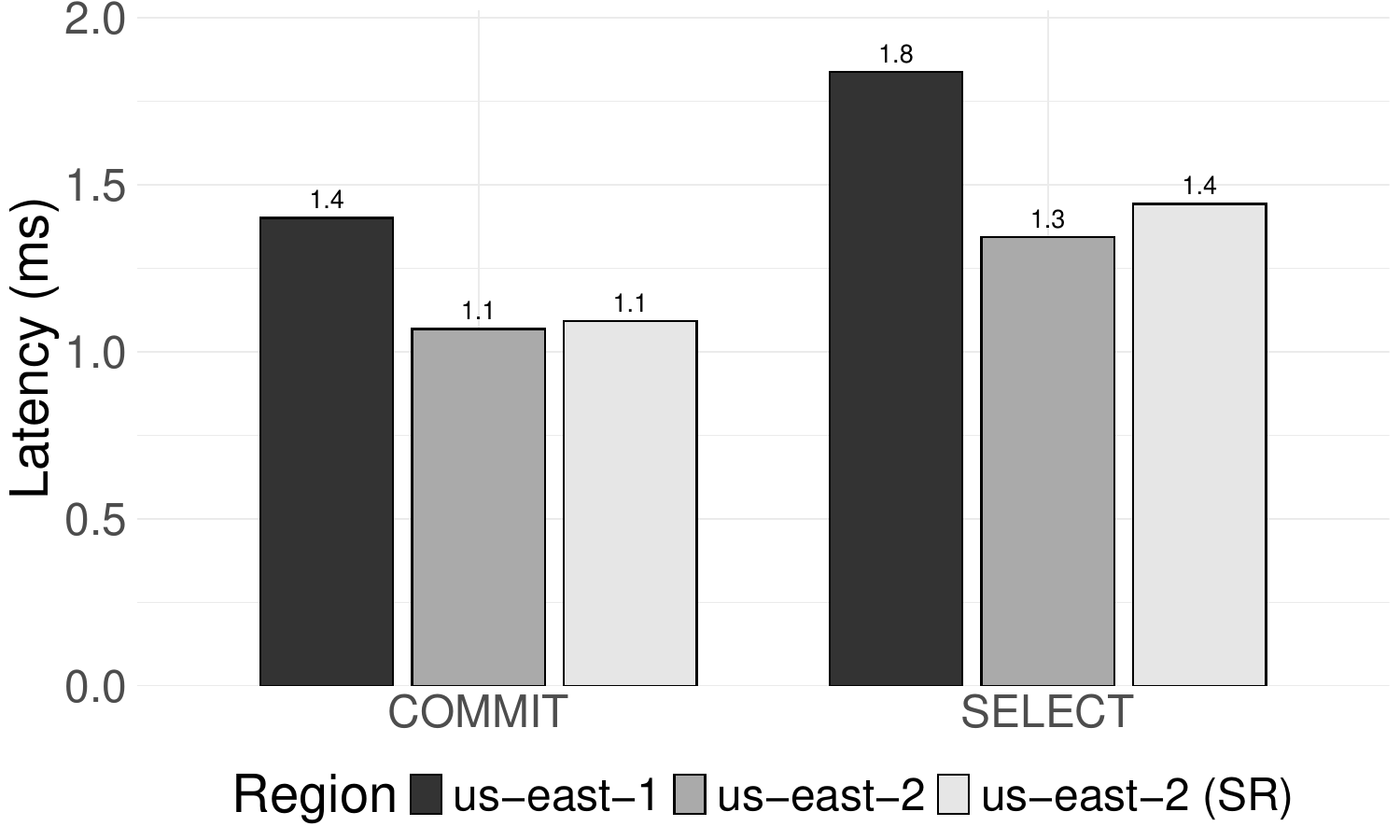}
\caption{DSQL 99th percentile latency for \texttt{COMMIT}, and \texttt{SELECT} operations for a two-region DSQL setup, and a single-region setup for a read-only transaction.}
\label{fig:low_level_latency_ro}
\end{figure}

\autoref{fig:low_level_latency_ro} shows the latency of \texttt{COMMIT} and primary key \texttt{SELECT} for a read-only transaction, for the same setup as \autoref{fig:low_level_latency}. Here we see that reads remain fast and local (in-region, and even in-AZ) even for the multi-region setups. For these read-only transactions, \texttt{COMMIT} is a no-op (discussed in \autoref{sec:reads}), and is handled locally by the QP. In this benchmark we use an explicit \texttt{COMMIT}, with latency $< 1.5ms$ p99.

Applications don't need to worry about whether they are single- or multi-region at all for read-only workloads, and experience identical performance. For read-write workloads, the only change is that the \texttt{COMMIT} will take additional time based on the region setup (approximately 2x the round-trip time between the nearest pair of regions).

These microbenchmarks use simple primary-key transactions to illustrate the low-level performance characteristics of multi-region DSQL\footnote{A full description of the benchmark, along with the benchmark code, are available at \url{https://github.com/mbrooker/...}}.

\subsection{Comparisons To Other Systems, Briefly}

DSQL's architecture is probably most similar to FoundationDB~\cite{zhou2021}, although with a significantly different approach to logging, atomic commitment, and replication. DSQL also has a scalable SQL engine (see \autoref{sec:sql-engine}), predicate pushdown to storage, multi-region, and supports interactive transactions. Aurora\cite{verbitski2017,verbitski2018}, Neon, AlloyDB and similar systems use distributed storage, but have a single writable leader. Layers like Vitess~\cite{Sougoumarane2012} and Citus~\cite{Umur2021} scale out these single-leader systems by partitioning the key space over leaders, typically sacrificing transaction atomicity and isolation. CockroachDB~\cite{Taft2020} and Spanner~\cite{Corbett2012} take a similar distributed SQL approach, but use pessimistic concurrency control, have a single leader per write shard with an associated \emph{lock table}, and replicate using Paxos groups rather than a disaggregated Journal. DSQL and Spanner are alike in their use of physical clocks, which CockroachDB uses Hybrid Logical Clocks~\cite{Kimball2022}. Once a transaction's \texttt{COMMIT} starts, DSQL's approach has a lot of similarities to the deterministic order-based approach of Calvin~\cite{thomson2012} and Slog~\cite{Ren2019}. DynamoDB~\cite{Elhemali2022} is a transactional~\cite{Idziorek2023} NoSQL system, with similar serverless properties, and support for multi-region applications. Architecturally, DynamoDB has more in common with Spanner with its use of sharding over Paxos groups, but uses an approach to transactions based on timestamp ordering, not unlike FoundationDB's. Like DSQL, MemoryDB~\cite{taleb2024} uses a disaggregated Journal service for replication, but has a single writable leader and eventually consistent scale-out reads. 

\section{Hosting the SQL Engine}
\label{sec:sql-engine}

To make it easier to achieve our familiarity goal, and to take advantage of decades of high-quality implementation work, DSQL uses the PostgreSQL engine for SQL parsing, execution, and optimization, as well as for the core implementation of the PostgreSQL protocol. We also wanted to offer strong security isolation between customers and between sessions, and the ability to scale SQL processing, compute, and connections out without the bounds of a single machine.

Each transaction in DSQL runs within a customized PostgreSQL engine hosted inside a dedicated Firecracker~\cite{agache2020} MicroVM. This is similar to PostgreSQL's process-per-session model, with the additional security protection of wrapping each process in a hardware-secured virtual machine. These MicroVMs are provisioned dynamically based on workload demands, with the system automatically scaling up by adding MicroVMs in the availability zones and regions where client connections originate. This approach ensures that SQL query processors remain geographically close to clients, optimizing for latency in the inherently chatty client-database interaction pattern typical of SQL workloads.

Creating a new MicroVM for a cluster doesn't require booting a virtual machine and starting PostgreSQL. Instead, the OS and database engine are started once, and a snapshot of the memory, register, and device state are taken. This snapshot is then cloned and restored when needed. This approach reduces restore time substantially, and also allows unmodified memory pages (such as those containing the kernel and engine code) to be shared across QPs for the same cluster with copy-on-write (much as they would be shared between processes in PostgreSQL's default \texttt{fork()}-based model).

The choice of PostgreSQL as the SQL engine foundation was driven by its proven pedigree, modularity, extensibility, and performance characteristics. However, DSQL uses only specific components of PostgreSQL—the SQL engine, an adapted version of the query planner and optimizer, and the client protocol implementation. We don't use PostgreSQL's storage or transaction processing components, instead using the distributed transaction and storage layers we describe in the next sections. DSQL bypasses the traditional buffer pool by plugging in at PostgreSQL's Access Method (AM) layer. Our storage plugin implements the AM interface directly, fetching data from DSQL's distributed storage without going through shared buffers or the lock manager. This eliminates buffer pool contention and lock manager overhead while maintaining compatibility with the query executor above the AM interface.

Each DSQL query processor operates as an independent unit that never communicates directly with other query processors. This shared-nothing architecture eliminates the coordination overhead typically associated with distributed SQL systems while still maintaining ACID transaction guarantees through the underlying storage and transaction management layers. Each QP executes one transaction at a time, with statements executing sequentially. Within a statement, reads fan out to multiple storage shards in parallel, and at commit time the prepare phase is broadcast to all involved adjudicators concurrently. When clients connect to DSQL, the system ensures sufficient MicroVM capacity exists to serve the load, dynamically provisioning additional compute resources as needed. Because query processors hold no durable state (writes are buffered locally only until commit), a QP failure simply aborts the in-flight transaction, which the client can retry on a freshly provisioned MicroVM.

In front of the MicroVMs we run a proxy (conceptually similar to the popular \texttt{pgbouncer}), dedicated to each database, which terminates the PostgreSQL wire protocol, and performs basic authentication and authorization. As sessions start a transaction (e.g. with \texttt{BEGIN}), they are either matched to an existing MicroVM, or a new one is created for the transactions. When transactions end, their MicroVMs are returned to the (dedicated per-database) pool for re-use by later transactions, or destroyed if the system detects excess capacity.

\section{Handling Reads}
\label{sec:reads}

Aurora DSQL's read path is designed around providing strongly consistent, snapshot-isolated access to data without requiring coordination between query processors or storage replicas. The system achieves this through a combination of precision timing and multiversion concurrency control (MVCC).

Every transaction in DSQL begins by selecting a transaction start time, $\tau_{start}$, using EC2's precision time infrastructure, which provides microsecond-accurate clocks with strong error bounds across all AWS regions. This timestamp serves as the foundation for snapshot isolation. All reads within the transaction request data \emph{as of} $\tau_{start}$, ensuring a point-in-time consistent view of the database regardless of concurrent writes or the specific storage shards and replicas accessed.

The storage layer implements these temporal reads using multiversion concurrency control, maintaining multiple versions of each row to support access to historical states without blocking concurrent operations. When a query processor requests data as of $\tau_{start}$, the storage system returns the most recent version of each row that was committed before that timestamp. This approach ensures that transactions see all data committed before $\tau_{start}$, no data committed after $\tau_{start}$, and no in-flight transactions. If the storage node is not current up to $\tau_{start}$, the reader is made to wait until replication catches up.

Read operations never require communication with a primary replica or lock server for sequencing, as they maintain no lock state. Reads can be served from the nearest replica within the same region and availability zone, minimizing both latency and cross-AZ data transfer costs. The system supports unlimited read replicas without coordination overhead, and readers never block writers or other readers.

The read path optimization extends to both read-only and read-write transactions, and does not require clients to declare transactions as read-only to get the full latency and scalability benefit.

DSQL storage nodes contain a logical copy of the data, with knowledge of the data schema: they store rows and index entries rather than opaque physical pages (in contrast with Aurora, which stores physical pages in its distributed storage layer~\cite{verbitski2017}). Each storage node stores only its shard of the data, not the entire dataset. This logical storage interface enables query pushdown capabilities that further optimize read performance. Rather than requesting pages, query processors request rows and delegate operations like filtering, aggregation, and projection to storage replicas. The storage system can perform index-only scans and complex filtering operations directly on the data, leveraging the principle that moving computation to data is more efficient than moving data to computation.

The absence of large coherent caches in the compute layer represents another architectural choice that enhances scalability. By avoiding cache coherence protocols and the associated coordination overhead, DSQL can scale compute resources independently. Some slowly-changing but frequently-read data, such as the catalog, is cached inside every query processor, avoiding round trips. Coherence of this smaller cache is provided by the transaction commit protocol. By comparison, Aurora PostgreSQL relies heavily on a large local cache on the single writer for performance, a trait shared by most similar single-writer database designs (both those with disaggregated storage and with local storage). The lack of cache coherence in these designs is exposed to clients as eventual consistency in exchange for read scale (because read replicas have their own cache, not coherent with the writers). Coherent cache designs like ScaleStore~\cite{ziegler2023,Ziegler2022} its precursors (including Rdb/VMS~\cite{lomet1992} in the early 1990s) require tightly coupled clusters which limit scalability, and fault tolerance beyond single datacenters.

For read-only transactions a \texttt{COMMIT} is a no-op. The Query Processor simply forgets about the transaction. No releasing of locks, or further communication with storage servers is needed. Read-only transactions never abort. If a storage node fails, in-flight reads to that node are retried against another replica of the same shard. A replacement node is recreated from S3 and the Journal (as described in \autoref{sec:writes}), and can begin serving reads once it has caught up to the current Journal position.

\subsection{Shard Scheme}\label{sec:shard_scheme}
In a sharded database, there are broadly two ways to spread data over multiple shards: ranges, or hashing. Hash partitioning, as used in DynamoDB, Dynamo~\cite{decandia2007} and many others, intentionally destroys spatial locality in the data, effectively spreading heat out over the key space, but losing optimizations based on locality (notably making in-order scans much chattier). Range-based partitioning, on the other hand, preserves spatial locality and offers superior performance on access patterns that take advantage of that, at the cost of a higher chance of hot-spotting. In DSQL, our partitioning scheme is range based, based on the observation that many OLTP SQL workloads have significant spatial locality. 

\section{Handling Writes and Commits}
\label{sec:writes}

In addition to using a consistent snapshot, read-write transactions need to check for conflicts to ensure snapshot isolation. This requires coordination. They also need to replicate changes for durability, and ensure correct transaction ordering.

Let's walk through what happens when you execute a write transaction. Consider this example:

\begin{small}
\begin{verbatim}
START TRANSACTION;
SELECT name, id FROM dogs;
UPDATE dogs SET latest_treat = now(), 
  treats = treats + 1 WHERE id = 5;
COMMIT;
\end{verbatim}
\end{small}

As described in \autoref{sec:reads}, we pick a transaction start time $\tau_{start}$ and perform all reads at that timestamp against our MVCC storage system. When the \texttt{UPDATE} statement executes, it doesn't write anything to storage yet. Instead, it records the planned change locally inside the Query Processor running this transaction. This keeps the \texttt{UPDATE} fast, in-region, and in the same availability zone as the application.

On \texttt{COMMIT}, three critical things need to happen: first, check whether the isolation rules allow the transaction to commit (or if it conflicts with other concurrent transactions and must abort); second, make the transaction results durable and atomic; and third, replicate the transaction to all AZs and regions where it needs to be visible.

The commit protocol starts by picking a set of adjudicators involved in the transaction (in our example only a single adjudicator will be involved, given that only one row is written). A protocol across these adjudicators picks a commit time $\tau_{commit}$ for the transaction. To achieve snapshot isolation, we need to detect write-write conflicts, whether any other transaction wrote the same keys between $\tau_{start}$ and $\tau_{commit}$. We don't worry about transactions that committed before $\tau_{start}$ because we've already seen their effects, and we don't worry about transactions committing after $\tau_{commit}$ because we don't see their effects (though they might not be able to commit due to our changes).

If no write-write conflicts are detected, one of the involved adjudictors will write the transaction to its Journal, in the form of a logical post image. Journal ensures that this write is atomic, and that no prior transactions have been committed to this Journal with a time stamp later than $\tau_{commit}$ (ensuring a per-Journal total transaction order). This post-image is then read from the Journal by the crossbar, and distributed to storage nodes which own a relevant portion of the key space.

The Journal is a commit log, with a payload of ordered, committed transactions that can be applied deterministically (and even without local read-modify-write on the storage node). This means storage replicas don't need any coordination when consuming the journal. In this sense, the post-commit portion of DSQL is somewhat similar to deterministic databases like Calvin~\cite{thomson2012} or SLOG~\cite{Ren2019}. Unlike these systems, however, DSQL can execute arbitrary interactive SQL transactions. DSQL's approach to replication is significantly different to DynamoDB~\cite{Elhemali2022}, Spanner~\cite{Corbett2012}, or CockroachDB~\cite{Taft2020}, all of which use Paxos variants for replication within a fixed group of replicas of each shard. DSQL's approach adds another layer, but allows for any number of read replicas of a shard, overlapping shard key spaces, and other flexibility.

The Journal is also continuously consumed to create a snapshot of the database state in S3. The combination of the durable copy of recent changes in the Journal, and this complete copy of older data in S3, means that DSQL storage is not involved in durability. This allows for the significant optimization that storage nodes don't need to sync to disk. They can use local SSDs to spill data that is too large for memory, but don't need to ensure that data is durable. If a storage node fails, or a new one is needed for a new read replica or shard, it is recreated from S3 and the Journal.

The consistency story has one more crucial piece: storage systems need to know they've seen all transactions with $\tau_{commit} \leq \tau_{start}$ before they can serve reads at $\tau_{start}$. The adjudicator handles this by promising never to commit transactions at earlier timestamps once it's committed at $\tau_{commit}$. We use a heartbeat protocol where adjudicators move their commit points forward in lockstep with the physical clock, ensuring storage knows when it has complete data.

\subsection{Detecting Conflicts, More Formally}

Starting with a commit that goes to a single adjudicator, let's formalize what the adjudicator does:

Committing transaction $A$, with start time $\tau^A_{start}$ and commit time $\tau^A_{commit}$:

\begin{enumerate}
\item Calculate $W_A$, the set of keys written by $A$.
\item Calculate $W_C$, the set of potentially conflicting keys, where $W_C$ is the union of $W_t$ for all committed transactions $t$ with $\tau^A_{start} < \tau^t_{commit} < \tau^A_{commit}$
\item If $W_A \cap W_C = \emptyset$ then $A$ can commit\label{step_noconflict}.
\item Promise to commit no additional transactions B with $W_A \cap W_B \neq \emptyset$ and $\tau^B_{commit} \leq \tau^A_{commit}$.
\end{enumerate}

In the case of commits where $W_A$'s keys span multiple adjudicators, step \ref{step_noconflict} becomes a \emph{yes} vote in the multi-adjudicator protocol. If all $W_A$'s keys belong to a single adjudicator, then that adjudicator can go ahead and commit the transaction to its Journal (by writing post-images of all the modified rows).

\subsubsection{Committing Across Multiple Adjudicators}
At scale, transactions will perform writes which span multiple adjudicators. Our cross-adjudicator commit protocol is a two-phase commit (2PC) variant with some inspiration from Warp~\cite{Escriva2015}. It proceeds as follows, given the set of Adjudicators $A$ involved in the transaction: 

\begin{enumerate}
\item one adjudicator $a_l$ is chosen from $A$,
\item the prepare phase is broadcasted by the query processor to all $a \in A$, each of which forwards their votes and current times $\tau_a$ to $a_l$ along with a promise not to commit any conflicting transactions with $\tau_{commit} < \tau_{max}$ (where $\tau_{max} = \tau_{commit}$ plus a system-chosen timeout),
\item once $a_l$ has received all positive votes, it calculates $\tau_{commit}$ as the max of all $\tau_a$,
\item if $\tau_{commit}$ exceeds $\mathrm{min(\tau_{max})}$ the transaction is abandoned,
\item otherwise it is written to $a_l$'s Journal atomically at time $\tau_{commit}$, a \emph{go ahead} message is broadcasted to $A$, and the query processor and client told the good news.
\end{enumerate}

If the voting phase takes too long, an adjudicator fails, or the \emph{go ahead} broadcast is not delivered, the other involved adjudicators will continue to process other conflicting transactions after the timeout $T$. Notably, this protocol is not an atomic commitment at all, but rather just an atomic voting on candidacy for commitment. $a_l$ can abandon the commit at any time before writing to the Journal without affecting correctness. The actual atomicity of the commitment is handled by the Journal, which accepts the entire transaction as a single atomic write. Because only $a_l$'s Journal is written to, there is no need for cross-Journal coordination or a fault-tolerant coordinator, avoiding the pitfalls of fault-tolerant atomic commitment generally~\cite{gray2006}. The other adjudicators in $A$ do not write to their own Journals for this transaction; their role is limited to voting on conflicts and temporarily holding promises. The crossbar then distributes the committed transaction from $a_l$'s Journal to all relevant storage shards, regardless of which adjudicators were involved.

Failed adjudicators are replaced by standby adjudicators via a leader election protocol. This can happen quickly due to the small amount of critical state that adjudicators require. In-flight transactions that involve a failed adjudicator are aborted: if the failed adjudicator is $a_l$, it has either already written to the Journal (in which case the transaction is committed and durable) or it has not (in which case the transaction is safely abandoned). If a non-leader adjudicator fails, $a_l$ will not receive its vote and will abandon the commit. In both cases, the promises held by surviving adjudicators expire after $\tau_{max}$, and the query processor can transparently retry the aborted transaction.

The protocol is correct for any arbitrary choice of $a_l$, but judicious choice of $a_l$ reduces the probability of deadlock. Nevertheless, deadlock can occur during the execution of this algorithm. In this case the later transaction is aborted, but its commit can be transparently retried. Due to the very short execution time of the commit algorithm (it only runs during commit, not during the transaction itself) this case of deadlock is extremely rare in the real-world workloads we have evaluated. There is no hard limit on the number of adjudicator shards a transaction may span, but commit latency grows with the number of involved adjudicators due to the fan-out of the prepare phase and the increased probability that $\tau_{commit}$ will exceed $\mathrm{min}(\tau_{max})$. In practice, OLTP transactions rarely span more than a handful of adjudicator shards.

\subsection{Index Maintenance}
The change written to the Journal isn't only a copy of the modified rows. It is also a post-image of any changes needed to maintain indexes on the modified data. Going back to our example, if there were an index on \texttt{latest\_treat}, the change written to the Journal would contain an additional update which removed \texttt{id = 5} from the previous value, and added it to the new value. These index changes are committed atomically alongside the modified data at the same $\tau_{commit}$, ensuring correct transactional and consistent behavior of indexes.

The calculation of index changes is performed, using the current catalog, by the Query Processor before submitting the candidate transaction to the adjudicator. The adjudicator treats these index changes the same way it treats regular updates, including for detecting conflicts. Two transactions which attempt to insert the same key into the same unique index would conflict, with the latter one being rejected.

While index updates are writes, we can optimize away the adjudicator check for conflicts on those written rows in many cases. This \emph{blind write optimization} is based on the common structure of secondary indexes, which are essentially inserting or removing a row's primary key from a set. Given that we're already checking for write-write conflicts on the primary key, conflicting set operations are prevented and the remaining operations are logically monotonic and can be safely issued in parallel~\cite{Hellerstein2020}. Depending on the table schema, some \texttt{INSERT}s can be optimized the same way. This optimization applies to all non-unique secondary indexes, including partial indexes and indexes with included columns, because their entries are keyed in part by the primary key and therefore cannot conflict independently of the base row. Unique indexes are excluded from the optimization: their conflict detection is handled through the normal adjudicator path, since distinct primary keys can produce duplicate unique index keys.

\subsection{Garbage Collection}
A common challenge in MVCC databases is garbage collection, removing old versions of rows that are no longer needed in the system. If these rows are removed too late, additional storage is needed, and caches are diluted, leading to lower performance. If garbage is collected too early, transactions that depend on the removed data need to abort. This problem is particularly acute in the DSQL architecture, because no part of the system knows about the full list of running transactions (in contrast to, say, PostgreSQL which keeps track of in-flight transactions).

To work around this issue, we chose to cap transaction run time. This allows simple, independent, time-based garbage collection to occur at each storage server. Servers can forget rows (other than the most recent) simply when they're older than a fixed time ($\tau_{expiry}$). If a read comes in with $\tau_{start} < \tau_{expiry}$, the storage node rejects it, and the query processor fails the transaction. While we don't believe this approach would be acceptable in analytics or OLAP systems, few OLTP workloads contain transactions that run for unbounded time. In our current deployment $\tau_{expiry}$ is set to five minutes before the current wall-clock time. To meet the needs of analytics use-cases, DSQL offers change data capture streams that allow easy ingestion into common analytics and warehousing systems (including Redshift).

\subsection{The Journal and Erasure Coding}
As mentioned above, Journal is a service we have been using internally at AWS for over a decade in various forms, including in processing many millions of transactions per second for S3, DynamoDB, and MemoryDB. While space does not permit diving deep into the implementation of Journal, we will mention that it uses a variant of chain replication~\cite{Renesse2004} for single-region replication (where chain replications simplicity and efficiency provide significant benefits), and a variant of Paxos for cross-region replication (where a quorum protocols ability to pick 2-of-3 latency is most important).

\begin{figure}[htbp]
\centering
\includegraphics[width=0.9\columnwidth]{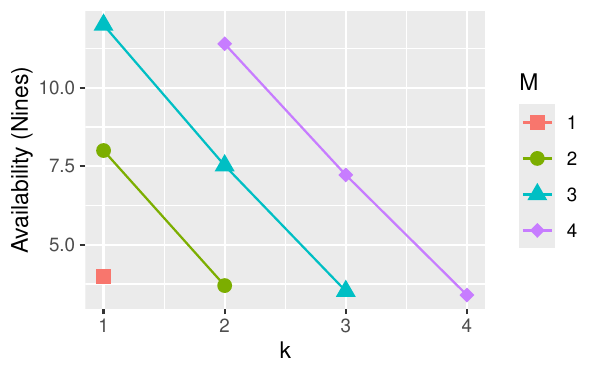}
\caption{Availability impact of erasure coding for various values of k and M, for base availability of 99.99\%}
\label{fig:ec_avail}
\end{figure}

To further reduce latency variance, we also don't use the Journal directly, but rather erasure code data across multiple Journals. This erasure coding is a latency optimization, although it also strictly improves durability, and significantly improves availability. A DSQL storage replica can't make progres on replication (and therefore can't serve reads) unless all the Journals it consumes from are available for reads. While Journal offers excellent availability, erasure coding across multiple Journals allows us to increase availability by multiple orders of magnitude at a fairly modest cost.

DSQL does a 2-of-3 code (i.e. erasure codes across three Journals, such that the data can be retrieved from any two). As shown in \autoref{fig:ec_avail}, for a base availability of 99.99\%, this $k=2$, $M=3$ code increases availability to beyond 7 and a half nines. While we considered multiple values of $k$ and $M$, this simple case met our availability goals, and has the additional benefit that the erasure code for $M=k+1$ is a trivial XOR. Erasure coding for availability is a widely used technique, but we believe that erasure coding for latency is an approach that should be more widely used by database and systems builders. If an individual Journal becomes unavailable, the erasure coding allows storage nodes to continue consuming committed transactions from the remaining Journals without interruption, and the commit path can continue writing to the available Journals while the failed one recovers.

\subsection{Exceptions to Snapshot Isolation}\label{sec:exceptions}
While snapshot isolation is nicely defined in the academic literature, these definitions don't map cleanly to the semantics that developers expect from SQL. One notable exception is the behavior of the catalog (the definition of tables and their schemas). DSQL, like PostgreSQL, stores the catalog in the database itself. The vast majority of OLTP transactions don't modify the catalog, and under the definitions of snapshot isolation would not be required to conflict with transactions (such as those doing \texttt{ALTER TABLE}) that do. However, this is insufficient. Consider the case of an \texttt{INSERT} transaction concurrent with an \texttt{ALTER TABLE}: the \texttt{ALTER} may modify the table, between the \texttt{UPDATE}'s $\tau_{start}$ and $\tau_{commit}$, in a way that makes the \texttt{UPDATE} invalid or illegal. To fix this problem, we additionally detect read-write conflicts on the catalog, making catalog table updates always serializable.

A similar case is explicit locking clauses, such as \texttt{FOR UPDATE}. These need to be treated specially, depending on the requested lock strength. In the case of \texttt{FOR UPDATE} we detect read-write conflicts on these rows, aborting the transaction if they have been modified since $\tau_{start}$. These are not the only exceptions: SQL is full of such edge cases. In general, academic definitions of snapshot isolation do not account for DDL or explicit locking, creating the need for selectively stronger isolation on specific operations. DSQL handles these cases within the existing OCC framework by extending the conflict check to include read-write conflicts on the affected rows, without changing the default isolation level for ordinary DML.

\subsection{Effects of Clock Skew}
The transaction start time $\tau_{start}$ is picked directly from the query processor's local clock, and $\tau_{commit}$ is picked by combining the local clocks and highest observed sequence numbers of all the adjudicators involved in a transaction. These local clocks are very high quality, with typical skew significantly below one network round-trip time, but we still need to consider what happens when physical clocks are inaccurate.

Let's call the \emph{true} physical time $\tau^p$. If $\tau_{start} \gg \tau^{p}_{start}$, the transactions reads will be delayed until the stream of Journal heartbeats catches up to $\tau_{start}$, affecting latency but not correctness. If $\tau_{start} \ll \tau^p_{start}$, the client may be able to observe a snapshot from before reads it knew to be committed, violating linearizability but not isolation. If $\tau_{commit} \gg \tau^p_{commit}$, then future transactions with $\tau_{start} = \tau^p_{start}$ may not observe the effects of this committed transaction, violating linearizability. Future transactions on the same adjudicator can't move time backwards, and so will need to block until $\tau^p$ catches up with the highest committed transaction time, affecting latency. $\tau_{commit} \ll \tau^p_{commit}$ will affect liveness, because storage will need to catch up to $\tau_p$ before processing any reads. Our clock hardware provides us with error bounds on the true clock time, and the correct edge is chosen in each case to preserve correctness.

In short, clock skew beyond the expected bounds causes the system to lose linearizability, but not isolation, durability, or atomicity. In other words, DSQL becomes merely snapshot isolated, rather than strong snapshot isolated. To avoid this condition, we closely monitor clock skew across our fleet using multiple approaches, and our clock synchronization hardware has a high level of internal redundancy. We believe that strong consistency is very important to application programmer's ability to write correct business logic, and treat any clock skew as a failure.

\section{Testing and Correctness}\label{sec:testing}

Correctness is, along with durability, the most important property that customers expect from database systems. While building our query processor on PostgreSQL gave us a significant head-start on correctness, replacing the storage layer, concurrency control, and other key components required an extensive approach to testing and validation.

Our approach to correctness starts with formal verification of the key protocols, an approach we have used extensively at AWS~\cite{newcombe2015, brooker2025}. We specified the core protocols in TLA+ and P, and performed extensive model checking. During this phase, we also explored protocol optimizations, with model checking giving us confidence that the proposed optimizations were correct.

We test the implementation extensively at build time using deterministic simulation testing. This approach (pioneered by FoundationDB~\cite{zhou2021}) requires modifying the code to be fully deterministic (notably running inside a deterministic thread scheduler), and building a simulation framework that can simulate behaviors like loss, latency, and re-ordering in the network. We developed \emph{turmoil}, a framework in Rust for this purpose. Deterministic simulation testing allows us to write build-time tests which test system behaviors, getting much better test coverage than is possible when the system is running on real infrastructure, by making tests cheaper and faster to run. While we do test the happy-case path during simulation testing, the focus is on the system's ability to remain correct while handling errors and failures. Our experience, and data from Yuan et al~\cite{yuan2014}, show that the majority of bugs in complex distributed systems are in error handling logic.

Once deployed, we test the system using fault injection testing while under load, validating that the failure handling results from simulation are correct. A subset of our in-production fault injection tests are available to DSQL customers via the AWS Fault Injection Service.

To validate that our SQL results match those returned by PostgreSQL, we deployed extensive fuzz testing alongside a static set of hundreds of fixed-function tests. The fuzz-testing approach, building on the approach of SQLancer~\cite{ba2023}, generates millions of example SQL statements and runs them both on DSQL and on Aurora PostgreSQL. In cases where our feature set fully overlaps with PostgreSQL, we expect DSQL to generate identical results to PostgreSQL. Some differences in behavior are tolerated where allowed by the SQL specification. For example in a \texttt{SELECT} without an explicit \texttt{ORDER BY} results may be returned in a different order, although the set of results must be identical. This approach has helped us find bugs in edge cases around floating point, collations, and NULL handling.

\begin{figure}[htbp]
\centering
\includegraphics[width=\columnwidth]{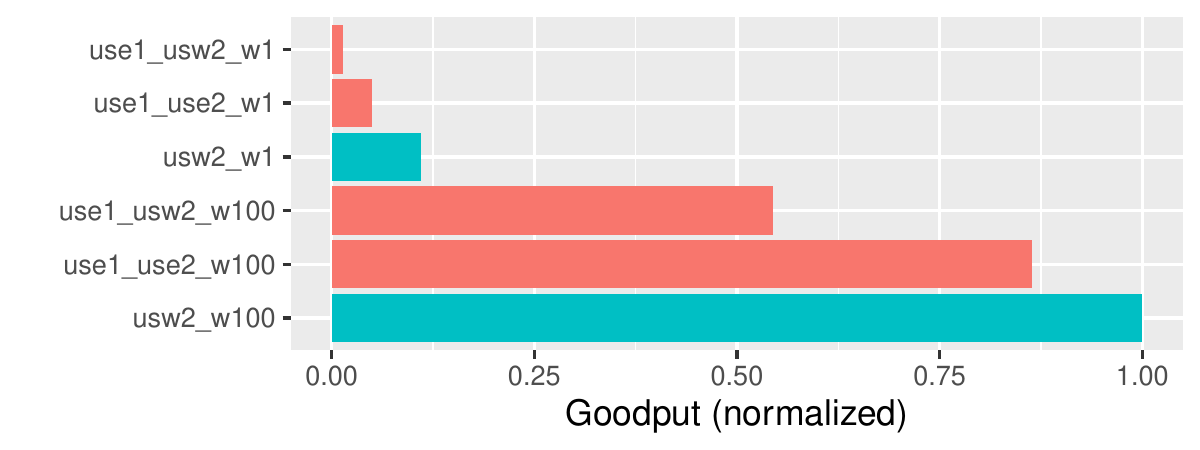}
\caption{Simulation results for a TPC-C-style benchmark.}
\label{fig:tpc-sim}
\end{figure}

Another pre-design validation was to build a event-based numerical simulation, which allowed us to explore the expected performance of the system under various workloads and design variations. As an example, \autoref{fig:tpc-sim} shows the results of one such simulation, comparing the expected goodput (throughput of committed transactions) for a TPC-C like benchmark run in three different region configurations (single region, multi-region between Virginia and Ohio, and multi-region between Virginia and Oregon). The limited concurrency nature of TPC-C-like benchmarks shows that the additional latency of cross-region commits can reduce workloads when client concurrency is limited, but that the overall system scales well.

Simulation results validated well against in-production testing. Numerical simulation allowed us to explore optimizations and design options quickly and confidently, without needing to build the whole system first.

\section{Control Plane and Heat Management}\label{sec:heat}

In Sections \ref{sec:reads} and \ref{sec:writes}, we described reads and writes being sent to the sharded storage servers and adjudicators, but not how these servers are found. When each query processor starts up, it is bootstrapped with a small per-database data set maintained by the control plane, which allows it to discover the storage servers and adjudicators for that database. The query processor then loads the catalog, which contains (much like PostgreSQL's catalog) the definition of tables and their schemas, statistics used for query optimization, and the authoritative partition map which maps the key space to adjudicators and storage servers.

The partition map must be kept up-to-date for liveness and availability, but using an incorrect partition map does not affect safety or correctness. An adjudicator knows for sure whether it's the leader of a given partition at a given $\tau_{commit}$, and a storage server knows for sure whether it's complete for a key or key range at a given $\tau_{start}$. Attempts to send requests to the wrong place are rejected.

\begin{figure}[htbp]
\centering
\includegraphics[width=0.6\columnwidth]{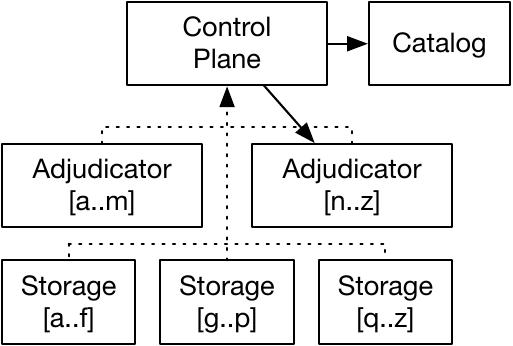}
\caption{Heat handling in the control plane}
\label{fig:cp_heat_overview}
\end{figure}

Maintaining the partition map is the control plane's most important job. When a new database is created, it'll contain a single storage partition (likely with three replicas, one per AZ), and a single adjudicator partition. As the database is used, the storage server and adjudicator send information about heat (read and write rate and throughput) and storage size across the key range to the control plane as illustrated in \autoref{fig:cp_heat_overview}. The control plane monitors this heat and size information, and uses a predictive approach to decide when and where storage servers or adjudicators need to be split. This is done significantly ahead of when resources are exhausted, to ensure availability.

The control plane plans and orchestrates splits. In the case of storage servers, this includes provisioning a new storage server, restoring data for the key range it owns from S3, subscribing to changes from the crossbar and adding this new server to the partition map. The system keeps storage shards small enough that new servers can serve traffic in seconds. Then the control plane instructs the previous server to clean up keys it no longer needs to track (and narrow its crossbar subscription). Because storage shards may temporarily have overlapping key ranges during this process, splits do not need to be atomic, significantly simplifying the scaling path. Splitting adjudicators is made slightly more complex by the tighter correctness requirements, namely that each key must be owned by at most one adjudicator while each key must be owned by at least one storage server, but simplified by the significantly smaller state.

Just like splits, the control plane can also choose to merge shards, and add replicas to any storage shard.

For storage servers:

\begin{description}
  \item[More shards] allows for more read and write throughput across the key space, along with more storage space for data.
  \item[Fewer shards] reduces the percentage of queries that must span multiple shards, improves spatial locality of data, with benefits for both performance and availability.
  \item[More replicas] allows for more read throughput for a given key (or small range of keys), and better locality of replicas to query processors.
  \item[Fewer replicas] reduces storage and write processing costs. Writes for a key range need to be applied to all replicas, making write cost scale with $O(\textrm{write rate} * \textrm{replicas})$.
\end{description}

The ability to control sharding and replication for storage independently was a key lesson we took from DynamoDB, which does not have the flexibility to handle hot read ranges through increasing the replication factor.

For adjudicators:

\begin{description}
  \item[More shards] allows for more write throughput across the key space.
  \item[Fewer shards] reduces the percentage of commits that need to execute the multi-adjudicator protocol, with benefits for commit performance and scalability.
\end{description}

The importance of the control plane and heat management to the overall performance of DSQL (and cloud services generally) should not be underestimated. The ability to monitor and predict heat distributions, and quickly and correctly orchestrate splits, merges, and replications is critical to the overall performance of the system.

\section{Lessons From Production}
\label{sec:lessons}
While we tried to learn as many lessons from AWS's existing databases as possible, getting the system into production has inevitably come with some key lessons, both operationally and from listening to customers.

\subsection{Commit Size Limits}
Currently, DSQL only allows any single transaction to modify a maximum of 3,000 rows, and 10MiB of data. We based the need for a transaction size limit on our operational experience at AWS around the importance of bounding tail latencies for stable applications. This operational necessity is partially driven by Little's Law~\cite{little1961proof}, which tells us that concurrency in a system in linearly proportional to the mean response latency. Concurrency, in turn, is the driver of resource allocation and transaction contention. Large writes in relational databases have a large impact on mean latency (much larger than that measured by limited-concurrency benchmarks like TPC-C, due to coordination omission~\cite{schroeder2006, Tene2012}). Transactions must consistently observe the world after or before, but never during, a transaction, causing head-of-line blocking of readers and potentially other writers. Capping transaction size to a size that can be applied in milliseconds limits this effect, leading to more stable, more predictable applications in keeping with our goal of simplifying application architectures.

However, limiting transaction size is also inconvenient for some applications, most notably during large data loads. We have heard more customer feedback than we expected about the difficulties of working through transaction size limitations, and we are working on a change to give customers who aren't interested in the benefits of consistent latency more control over transaction sizes.

\subsection{Foreign Key Constraints}
In the original production release of DSQL, we don't support enforcement of foreign key constraints (FKCs), although we do fully support \texttt{JOIN} and schemas with foreign key relationships. We chose this as a time-to-market optimization, given conversations with at-scale customers who don't use FKCs for performance reasons, and small customers who see little value in them. We have been surprised by more demand for FKCs than we expected, and are working to implement them.

As with the cases discussed in \autoref{sec:exceptions}, it is not possible to implement FKCs correctly under pure snapshot isolation. Our implementation of FKCs will introduce checking for read-write conflicts in the case of FK relationships, similar but not identical to the implementation of \texttt{FOR UPDATE}. The performance impact of this is to increase how often transactions will need to span multiple adjudicators, and run the more complex multi-adjudicator commit protocol. It is, however, not a significant architectural change.

\subsection{Sequences and Indexes}
We still believe we chose correctly when picking range partitioning (see \autoref{sec:shard_scheme}), there is three cases it makes significantly harder: supporting serial keys (aka sequences or \texttt{AUTO\_INCREMENT}), supporting secondary indexes with write locality (such as those that index on a \texttt{timestamp} column and always insert \texttt{now()}), and supporting indexes with low cardinality (such as indexes on \texttt{boolean}). In fact, we suspect that supporting these high locality cases at scale while preserving read locality may not be possible in general.

Our plan is to extend our range-based scheme to a hybrid scheme, allowing sub-ranges to be distributed hash-style over multiple storage nodes. This allows scaling of per-range write throughput, while still avoiding the hash key downside of ordered scans jumping around the whole storage fleet.

\subsection{Strong Consistency}
The 2007 Amazon Dynamo paper's embrace of eventual consistency~\cite{decandia2007}, and Werner Vogels 2009 article \emph{Eventually Consistent}~\cite{Vogels2009}, reflected the thinking at the time that cloud-scale systems needed to embrace eventual consistency to achieve their availability and latency goals. Since then, advancements in time distribution, datacenter networks, power and cooling infrastructure, and distributed protocols have changed the trade-offs. Two decades of working with application programmers, and building thousands of cloud-scale services at Amazon, have shown us the difficulties programmers face when trying to write correct business logic in the face of eventual consistency. In DSQL we have embraced strong consistency (namely linearizability), while still achieving nearly 10x lower read latency and 4x lower write latency than reported for 2007's Dynamo paper.

\section{Conclusion}
\label{sec:conclusion}

We believe that we have achieved our most important goal with Aurora DSQL: building a relational database that simplifies our customer's architectures, through offering strong semantics, scalability up and down, high availability and durability, and no operations. We have seen considerable success running production applications on DSQL. Overall, we believe that we have validated our hypothesis that a disaggregated SQL database built around a journal service can offer excellent features, performance, and operational properties. However, as with any project of this size, there is still work to do. We're working to add support for foreign key constraints, stored procedures, and other popular SQL features. We're also working actively on latency, throughput, and query planning and execution.

\begin{acks}
Like any cloud service, Aurora DSQL was the work of a talented and dedicated team, along with many partner organizations in AWS. We owe significant gratitude to all who contributed. In particular, we would like to thank Shyam Krishnamoorthy, Eric Kraemer, G2 Krishnamoorthy, Swami Sivasubramanian, Ankush Desai, Murat Demirbas, Gaurav Roy, Carlos Vara, Josh Levinson, Julien Ridoux, Rahul Pathak, and Yossi Levanoni. This work owes a significant debt to the earlier work of Al Vermeulen, and the teams at AWS that built JournalDB, QLDB, Aurora, and DynamoDB. 
\end{acks}


\bibliographystyle{ACM-Reference-Format}
\bibliography{paper}

\end{document}